\documentclass[reprint,amsmath,amssymb,aps,prx,longbibliography, balancelastpage,
floatfix,superscriptaddress,nofootinbib]{revtex4-2}
\usepackage[T1]{fontenc}
\usepackage{graphicx}
\usepackage{dcolumn}
\usepackage{bm}
\usepackage{mathtools}
\usepackage{hyperref}
\usepackage{xcolor}
\hypersetup{colorlinks=true,citecolor=[rgb]{0,0.15,0.60}, 
urlcolor=[rgb]{0,0.15,0.60}, linkcolor=[rgb]{0,0.15,0.60}, final = true}
\usepackage{physics}
\usepackage{times}
\usepackage{eucal}
\usepackage{relsize}

\usepackage{siunitx}
\DeclareSIUnit\molar{\mole\per\cubic\deci\metre}
\DeclareSIUnit\Molar{\textsc{m}}

\def\sb#1{\textbf{\textsf{#1}}}
\def\nsb#1{\noindent\textbf{\textsf{#1}}.~}

\setcitestyle{round,super}

\usepackage{type1cm}
\usepackage{lettrine}

\def\ie{{\it i.e.},~}

\def\r{\right}
\def\l{\left}
\def\lang{\left\langle}
\def\rang{\right\rangle}
\def\Rb{\mathbf{R}}

\def\eb{\mathbf{e}}
\def\mb{\mathbf{m}}
\def\rb{\mathbf{r}}
\def\vb{\mathbf{v}}

\def\ww{w}   
\def\xib{\boldsymbol{\xi}}
\def\half{{\textstyle\frac{1}{2}}}

\def\sA{{\mathsmaller{\mathrm{A}}}}
\def\sB{{\mathsmaller{\mathrm{B}}}}

\def\sH{{\mathsmaller{\mathrm{H}}}}

\def\sT{{\mathsmaller{\mathrm{T}}}}
\def\sW{{\mathsmaller{\mathrm{W}}}}
\def\sMFP{{\mathsmaller{\mathrm{MFP}}}}
\def\zT{\zeta_\sT}
\def\zA{\zeta_\sA}

\def\sigA{\sigma_\sA}
\def\sigT{\sigma_\sT}
\def\tauA{\tau_\sA}

\def\tauc{\tau^{\rm c}}
\def\AA{\mathcal{A}}
\def\FF{\mathcal{F}}
\def\GG{\mathcal{G}}
\def\NN{\mathcal{N}}
\def\RR{\mathcal{R}}
\def\bq{\bar{q}}
\def\bt{\bar{t}}

\def\bzT{\bar{\zT}}
\def\bzA{\bar{\zA}}
\def\bAA{\bar{\AA}}

\def\bzet{\bar{\zeta}}

\def\btau{\bar{\tau}}
\def\btauA{\btau_\sA}
\def\dbq{\dot{\bq}}
\def\kB{k_\sB}
\def\kBT{\kB T}
\def\YT{Y_\mathrm{T}}
\def\xiw{\xi_\sW}
\def\tauMFP{\tau_\sMFP}
\def\kappaT{\kappa_\sT}
\def\FH{F_\sH}
\def\EB{E_\sB}
\def\Ebeff{E_\sB^\mathrm{eff}}
\def\kapfast{\kappa_\mathrm{fast}}
\def\kapslow{\kappa_\mathrm{slow}}
\def\Fmax{F_\mathrm{max}}
\def\Pmax{P_\mathrm{max}}
\def\qmax{q_\mathrm{max}}
\def\Ueff{U^\mathrm{eff}}
\def\EBeff{\EB^\mathrm{eff}}
\def\source{{\mathsmaller{\mathrm{source}}}}
\def\target{{\mathsmaller{\mathrm{target}}}}
\def\tauFP{\tau_\mathsmaller{\mathrm{FP}}}

\definecolor{YKB}{rgb}{0.00,0.18,0.65}

\definecolor{NYL}{cmyk}{0.7,0.0,0.0,0.4}

\definecolor{GRY}{rgb}{0.5,0.5,0.5}

\begin{document}

\title{\sb{\larger Acceleration of enzymatic catalysis by active hydrodynamic fluctuations}}

\author{Ashwani Kr. Tripathi}
\affiliation{Center for Soft and Living Matter, Institute for Basic Science (IBS), Ulsan, 44919, Republic of Korea}
\author{Tamoghna Das}
\affiliation{Center for Soft and Living Matter, Institute for Basic Science (IBS), Ulsan, 44919, Republic of Korea}
\author{Govind Paneru} 
\affiliation{Center for Soft and Living Matter, Institute for Basic Science (IBS), Ulsan, 44919, Republic of Korea}
\author{Hyuk Kyu Pak}
\email{hyuk.k.pak@gmail.com}
\affiliation{Center for Soft and Living Matter, Institute for Basic Science (IBS), Ulsan, 44919, Republic of Korea}
\affiliation{Department of Physics, Ulsan National Institute of Science and Technology, Ulsan, 44919, Republic of Korea}
\author{Tsvi Tlusty}
\email{tsvitlusty@gmail.com}
\affiliation{Center for Soft and Living Matter, Institute for Basic Science (IBS), Ulsan, 44919, Republic of Korea}
\affiliation{Department of Physics, Ulsan National Institute of Science and Technology, Ulsan, 44919, Republic of Korea}
\affiliation{Department of Chemistry, Ulsan National Institute of Science and Technology, Ulsan, 44919, Republic of Korea}

\date{\today}

\maketitle
\onecolumngrid
 \noindent\textsf{\textbf{Abstract}\\
 The cellular milieu is teeming with biochemical nano-machines whose activity is a strong source of correlated non-thermal fluctuations termed active noise. Essential elements of this circuitry are enzymes, catalysts that speed up the rate of metabolic reactions by orders of magnitude, thereby making life possible. Here, we examine the possibility that active noise in the cell, or in vitro, affects enzymatic catalytic rate by accelerating or decelerating the crossing rate of energy barriers during the reaction. Considering hydrodynamic perturbations induced by biochemical activity as a source of active noise, we evaluate their impact on the enzymatic cycle using a combination of analytic and numerical methods. Our estimates show that the fast component of the active noise spectrum enhances the rate of enzymes, while reactions remain practically unaffected by the slow noise spectrum. Revisiting the physics of barrier crossing under the influence of active hydrodynamic fluctuations suggests that the biochemical activity of macromolecules such as enzymes is coupled to active noise. Thus, we propose that enzymatic catalysis is a collective, many-body process in which enzymes may affect each other's activity via long-range hydrodynamic interaction, with potential impact on biochemical networks in living and artificial systems alike.} \\
 \twocolumngrid

\date{\today}


\vspace{1.1cm}
\noindent The idea that enzymes achieve their phenomenal catalytic capacity by stabilizing an activated transition state was introduced by Haldane~\cite{Haldane1930} and developed by Pauling~\cite{Pauling1946} who lucidly stated this postulate:~\cite{Pauling1948}
``\ldots that the enzyme has a configuration complementary to the activated complex, and accordingly has the strongest power of attraction for the activated complex, means that the activation energy for the reaction is less in the presence of the enzyme than in its absence, and accordingly that the reaction would be speeded up by the enzyme.'' Electrostatic effects, chiefly the formation of a preorganized polar network, were recognized as pivotal in stabilizing the transition state.~\cite{Warshel1976,Warshel1978} In this extremely fruitful view of enzymatic catalysis, the activated complex is jolted past the transition state's energy barrier by thermal agitation.~\cite{Kraut1988,Fersht2017,Kessel2018} The cell, however, is bustling with activity that generates significant athermal agitation,~\cite{Guo2014,Turlier2016,Fodor2016,Battle2016,Ahmed2018} provoking the main question asked in this paper: 
how may athermal active noise affect enzymatic catalysis?  

During their catalytic cycle, many enzymes undergo conformational changes, for example to enable substrate binding and product release.~\cite{Austin1975,Gerstein1994,Hammes2002,Daniel2003, Gutteridge2005,Boehr2006a,Nagel2009,Glowacki2012, Bhabha2015, Callender2015,Palmer2015,Mitchell2016,Eckmann2019} Such internal motions and rearrangements are part of essential mechanisms, particularly induced fit,~\cite{Koshland1958} conformational selection,~\cite{Ma2010,Vertessy2011} allostery,~\cite{Monod1965,Perutz1970,Goodey2008,Motlagh2014, DuBay2015} and conformational proofreading.~\cite{Savir2007,Savir2010a,Savir2013} The coexistence of multiple conformational states~\cite{English2006} may assist evolution to explore new functions.~\cite{Campbell2016} Motor proteins operate by converting chemical energy into conformational changes and motion,~\cite{Howard1997, Vale2003, Kodera2010} and recent studies suggest that similar coupling underlies the boosted diffusion observed in the active enzymes.~\cite{Muddana2010, Dey2015, Zhao2017, Jee2018, Jee2018a}
Linkage between intrinsic motion and catalysis was reported in adenylate kinase (ADK),~\cite{WolfWatz2004, HenzlerWildman2007, HenzlerWildman2007a, Olsson2010,Aviram2018} dihydrofolate reductase (DHFR),~\cite{Schnell2004, Venkitakrishnan2004,  Boehr2006, HammesSchiffer2006, Bhabha2011, Luk2013, Hanoian2015} and other enzymes~\cite{Eisenmesser2005, HammesSchiffer2006, Kale2008}---though the existence, extent, and physical nature of this linkage remain open questions.~\cite{Nagel2009, Kamerlin2010, Glowacki2012}
All this invokes a notion of enzymes as stochastic molecular machines whose chemical performance and evolution are linked to their internal mechanics.~\cite{HammesSchiffer2006, Togashi2007,Flechsig2010, Hekstra2016, Ma2016, Dutta2018, Eckmann2019, Hosaka2020}

For their nanometric size, these machines are subject to violent, thermal and athermal, agitations by the fluctuating environment: Thermal white noise originates from memoryless equilibrium fluctuations. Athermal {\em colored} noise is generated by a variety of temporally-correlated active sources, such as molecular motors and cytoskeleton rearrangement,~\cite{Bursac2005, Fodor2016, Ahmed2018, Guo2014, BernheimGroswasser2018,Sens2020} and the dynamics of other cellular machinery, including enzymes.~\cite{Muddana2010, Dey2015, Zhao2017, Jee2018, Jee2018a}
This work lays out a simple model in order to investigate how these thermal and athermal fluctuations, in vivo or in vitro, might affect the catalytic reaction rate. From a coarse-grained perspective, we treat enzymes as stochastic force dipoles,~\cite{Manneville2001, Marchetti2013,  Mikhailov2015, Flechsig2019, Hosaka2020} whose internal motion represents conformational changes of the enzyme during the catalytic cycle. 

Transition state theory treats chemical reactions as thermal diffusion processes in energy landscapes whose coordinates capture the chemical transformation. Within this physical picture, thermal agitation drives the system from the initial stable state of reactants to the final stable state of the products along a stochastic pathway, crossing the energy barrier at a saddle point---a metastable transition state that governs the reaction rate.~\cite{Eyring1935, Kramers1940, Haenggi1990, Gammaitoni1998, Pollak2005}

Here, we extend the classical, thermally-induced transition state theory into an \emph{actively-induced transition state theory}, which accounts for the impact of correlated noise generated by hydrodynamic fluctuations.
The framework we developed allowed us to compute the reaction rate, relative to a purely thermally-fluctuating enzyme, as a function of the active noise strength and its correlation timescale. Within a biologically relevant parameter range typical to enzymes, we find two potential effects of active noise: Strong active noise with long correlation time (relative to the thermal turnover rate) hinders enzymatic activity, but not significantly. In contrast, active fluctuations of any strength with short to intermediate correlation times enhance the catalytic rate compared to a thermally-activated enzyme. Under the coaction of thermal and active forces, in a biologically relevant regime, we find a potential increase of about \SIrange{10}{180}{\percent} in the turnover rate of enzymes.
The present method is general and can be applied to other physical and biological processes that can be cast as an effective multi-state system with noisy memory, for example, unzipping of DNA and RNA hairpins~\cite{Woodside2006, Greenleaf2008, Woodside2008, Neupane2012, Vandebroek2017} or solutions of organic catalysts.~\cite{Wang2020}

\vspace{0.8cm}
\noindent\sb{\large Results and Discussion}\\

\noindent\nsb{Hydrodynamic fluctuations as active noise}
An enzyme in a cellular environment continually experiences correlated stochastic forces, as a collective effect of diverse flow-generating mechanisms, which we model as sources of athermal active noise. To estimate these stochastic forces, we approximate the active noise sources as an ensemble of force dipoles. This is a valid long-range approximation as force dipoles induce the leading term in the far-field expansion of momentum-conserving hydrodynamic perturbations.~\cite{Pozrikidis1992, Diamant2007}
Each force dipole is represented as two equal masses connected by a spring of rest length $\ell$. The cellular background is treated as a random ensemble of average concentration $c_0$ of such force dipoles whose moments $\{\mb_i\}$ are randomly distributed at positions $\{\Rb_i\}$ with randomly isotropic orientations $\{{\bf e}_i\}$. 

As the typical inertial timescale (${\sim}\SI{1}{\ps}$) is much shorter than the characteristic timescale of an enzyme, (${>}\SI{1}{\micro\s}$), the background flow is overdamped. It is therefore convenient to treat this linear Stokesian flow in terms of its Green function, the mobility tensor $\GG$. A dipole $\mb$ made of a pair of opposing point forces will therefore generate a flow field $\vb(\rb)$ proportional to the gradient of the Green function, $\vb(\rb) = \nabla \GG(\rb, \rb') \, \mb(\rb')$, where $\rb'$ is the position of the source dipole. A target dipole (\ie~an enzyme) of length $\ell_0$ subjected to this flow will experience an internal stress (tension or compression) proportional to the velocity gradient along its axis. This dipole-dipole force $\FH$ will therefore be proportional to the second derivatives of the mobility, $\FH \sim \eta \ww \ell_0 \nabla \vb \sim \eta \ww \ell_0 m \nabla\nabla \GG$, where $\ww$ is the hydrodynamic diameter of the dipole's beads and $\eta$ the viscosity (see  Methods for a detailed derivation). As biological flows are typically of low Reynolds number, $\GG$ can be approximated as the Oseen's tensor which scales $\GG \sim 1/(\eta r)$, where $r$ is dipole-dipole separation. Then, $\FH \sim \eta \ell_0^2 m \nabla\nabla \GG \sim \ell_0^2 m r^{-3}$ (taking $\ww \sim \ell_0$).

Since the force dipoles are randomly positioned and oriented, ensemble or time-averaging forbids the accumulation of net mean dipole moment, $\lang \mb_i(t)\rang = 0$. Thus, the average net flow and induced internal forces also vanish, $\lang \nabla v \rang \sim \lang \FH \rang = 0$. What survives averaging are of course the fluctuations experienced by the target dipole (the enzyme), $\lang ( \nabla v )^2 \rang \sim \langle \FH^2 \rangle \neq 0$. Summed over the random ensemble, the force fluctuations scale as
\begin{align}
    \label{eq:hydroforce}
    \lang \FH^2 \rang &\sim c_0 \lang m^2 \rang \ell_0^4
    \int_{\ell_0}^{\infty}{r^{-6} \dd^3r} \nonumber\\
&\sim c_0 \ell_0\lang m^2 \rang \sim \RR^{-3}\ell_0\lang m^2 \rang~,
\end{align}
where $\RR \equiv c_0^{-1/3}$ is the average dipole-dipole separation. The exact expression, derived in  Methods, includes a geometric factor of order unity. Eq.~[\ref{eq:hydroforce}] preserves the long-range nature of hydrodynamic fluctuations which decay as $c_0 \sim \RR^{-3}$. For typical concentrations of active sources, such as enzymes or motors, ranging between $c_0 \sim \SI{1}{\micro\Molar}-\SI{1}{\milli\Molar}$, $\RR\sim \SIrange{10}{100}{\nm}$. The dipole moment fluctuation can be approximated as $\lang m^2\rang\sim(\FF \ell)^2$, where $\ell$ the size of the active elements and $\FF$ is the net force they generate during their turnover cycle. For values typical to motor proteins, $\FF \sim \SI{1}{\pico\newton}$ and $\ell \sim \SI{5}{\nm}$,~\cite{Ahmed2018, Vale2003, Milo2015} the dipole fluctuations would be $\lang m^2\rang \sim \SI{1}{(\kB T)^2}$. We shall use the value of hydrodynamic fluctuation $\lang \FH^2 \rang$ as the strength of the active noise.

The sources of active noise in the cell have widely varied correlation timescales, and are interdependent components of an intertwined biochemical circuitry. However, the timescales of the network's collective dynamics are much longer than the correlation time of a single source. As suggested by recent experimental measurements,~\cite{Guo2014, Fodor2015, Fodor2016, Ahmed2018, Turlier2016} the sources might be assumed as independent stochastic processes with intermittent bursts of activity, each with its own auto-correlation statistics. Thus, we consider the background flow as an active noise $\zA(t)$ with a characteristic correlation time $\tauA$ realised as:
\begin{align}
    \label{eq:hydrocorrl}
    \lang \zA(t)\zA(t') \rang = \lang \FH^2 \rang \exp \l(-|t-t'|/\tauA \r)~.
\end{align}
Such activity maintains a certain type of fluctuation-dissipation relation, as observed in cells,~\cite{Fodor2015, Fodor2016,BernheimGroswasser2018} where injection (extraction) of an energy $\AA$ into (from) the system is compensated by the correlation time such that the noise strength, $\lang \FH^2 \rang=\AA/\tauA$ remains constant. With these considerations, we now proceed to derive the reaction rate theory in the presence of such active noise.\\

\nsb{Reaction in the presence of hydrodynamic fluctuations} 
We start by writing down the dynamical equation for a reaction occurring in an energy landscape under the influence of a noise $\zT(t)$ of thermal origin, and an active noise $\zA(t)$ resulting from the long-range correlated hydrodynamic fluctuations,
\begin{align}
    \label{eq:eom}
    \gamma\dot{q} &= F(q) + \zT(t) + \zA(t)~.
\end{align}

\begin{figure}[!b]
\includegraphics[width=0.95\columnwidth]{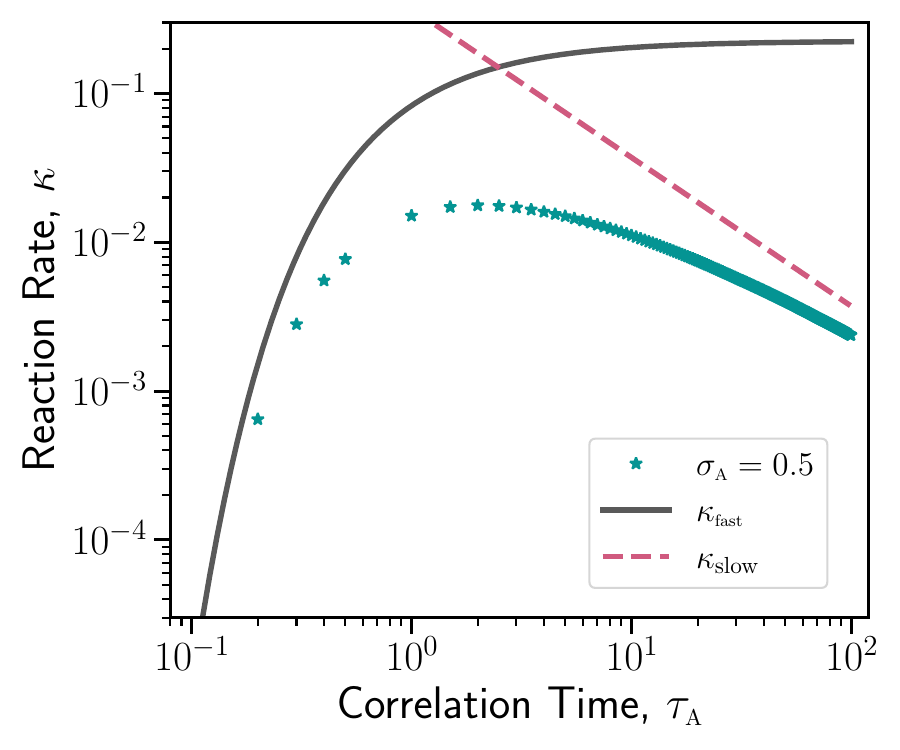}
\caption{\sb{Variation of reaction rate with noise correlation time.}
Data points show the reaction rate $\kappa$,
for the noise of strength $\sigA = 0.5$, for the case of pure active noise. $\kappa$ follows the asymptotic limits  $\kapslow$~(Eq.~[\ref{eq:slow}], dashed curve) and  $\kapfast,$~(Eq.~[\ref{eq:fast}], solid curve).
}
\label{fig:Act_Asy}
\end{figure}

The first term on the right hand side of Eq.~[\ref{eq:eom}] is a conservative force, $F(q)=-\partial_q U(q)$, exerted by the potential $U(q)=-(a/2) q^2 + (b/4) q^4$. The reaction potential $U(q)$ is made of two wells are positioned symmetrically at $q_0=\pm \sqrt{a/b}$, separated by an energy barrier $\EB=a^2/(4b)$ at $q=0$. $a, b > 0$ are  phenomenological constants. Physically, $a$ would represent the stiffness of a protein, roughly the spring constant of the force dipole and $b$ stands for the strength of the simplest possible anharmonicity that yields an activation barrier $\EB$. The internal friction of the landscape $\gamma$ sets the intrinsic timescale $\tau_0=\gamma/a$.
Note that Eq.~[\ref{eq:eom}] merely assumes that the catalytic cycle is amenable to stochastic active noise, as it is to thermal noise, but requires no coupling of reaction and conformational coordinates.

The thermal force $\zT(t)$ is drawn randomly from temporally uncorrelated white noise, $\langle \zT(t)\zT(t')\rangle = 2\gamma \kBT\delta(t-t')$ with the noise strength fixed by the temperature $T$, where $\kB$ is the Boltzmann constant. Unlike the thermal noise, the active noise $\zA(t)$ is temporally correlated (Eq.~[\ref{eq:hydrocorrl}]), which we ensure by modeling it with an Ornstein-Uhlenbeck type evolution dynamics:~\cite{Sharma2017}
\begin{align}
    \label{eq:activeevol}
    \tauA\dot{\zA} = -\zA+\sqrt{2\AA}\,\xiw(t)~,
\end{align}
where $\xiw(t)$ is a standard white noise process. Our main objective is to study the effect of active noise $\zA$ (with thermal noise $\zT$ in the background) on the reaction rate, $\kappa=1/(2\tauMFP)$, where $\tauMFP$ is the mean first passage time needed to cross the energy barrier.

\begin{figure*}[!htb]
\centering
\includegraphics[width=.99\textwidth]{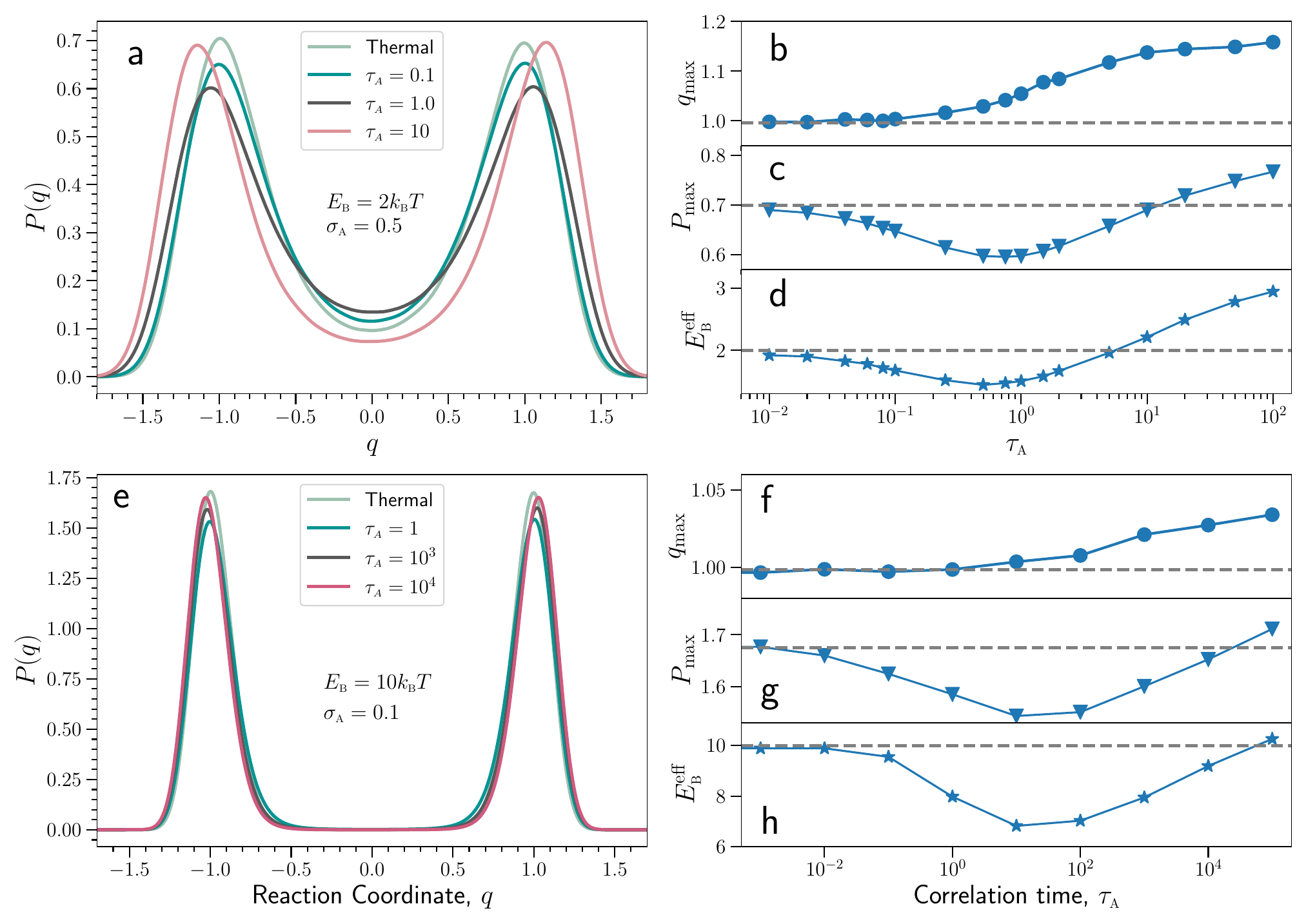}
\caption{\sb{Effect of active noise on the statistics of reaction dynamics.}
(\sb{a}-\sb{d}) and (\sb{e}-\sb{h}), represent the results for energy barriers, $\EB = 2 \kB T$ (\ie $\sigT = 0.5$) and $\EB = 10 \kB T$ (\ie $\sigT = 1/\sqrt{20}$), respectively.
(\sb{a}, \sb{e}) The probability density $P(q)$ in the reaction energy landscape. A purely thermal case~(Active noise strength, $\sigA=0$) is shown for reference. 
(\sb{b}, \sb{f}) The most probable position in the reaction landscape $\qmax$ is monotonically pushed away from the barrier, as compared to a purely thermal system whose maximum is at $q=q_0$ (dashed line).
(\sb{c}, \sb{g}) However, the maximal probability $\Pmax$ exhibits a non-monotonic dependence on $\tauA$. For very small $\tauA$, $\Pmax$ is lower than its expected thermal value (dashed line), and continues to decrease till $\tauA \sim 1, 50$ for $\EB=2, 10\kBT$, respectively, and then turns to increase and eventually becomes larger than the thermal value.
(\sb{d}, \sb{h}) Similar non-monotonic behavior of the effective reaction energy barrier $\Ebeff$, computed from the distribution $P(q)$. Note that, for $\tauA > \tauc$, $\Ebeff$ is larger than the purely thermal case~(dashed line).
}
\label{fig:probability}
\end{figure*}

To gain some intuition, we examine the asymptotic case of negligible thermal noise. Then, Eqs.~[\ref{eq:eom},\ref{eq:activeevol}], can be recast as an underdamped Langevin equation,
\begin{align}
    \label{eq:eom2}
    \l( \gamma \tauA \r) \ddot{q} = 
    -\Gamma(q)\,\dot{q} + F(q)+ \sqrt{2\AA}\,\xiw(t)~,
\end{align}
with the effective friction coefficient $\Gamma(q) \equiv \gamma -\tauA \,\partial_q F(q)$. An additional feature of this nonlinear dynamical equation is that the evolution of reaction depends, besides on the force $F(q)$ itself, also on its gradient, that is the  curvature of the potential, $\partial_q F = - \partial_q^2 U$. Thus, when $\tauA > \tau_0$, the effective friction $\Gamma(q)$ turns negative close to the energy barrier.~\cite{Caprini2019, Fily2019, Farage2015, Marconi2019} The negative friction region---where the motion is accelerated past the barrier---grows with $\tauA$, until it stretches between the two inflection points of the potential, $\abs{q}\leq q_0/\sqrt{3}$, for $\tauA\gg\tau_0$. The maximal force, $\Fmax = (2/\sqrt{3})^3 (\EB/q_0)$, would be experienced at these inflection points. In the long-memory regime, $\tauA\gg\tau_0$, any force $F\ge\Fmax$, is likely to push the reaction to the other potential well by crossing into the negative friction region. Thus, $\Fmax$ sets an effective  force barrier, similar to the energy barrier $\EB$ in the short-memory regime. As this phenomenology evidently affects the reaction rate, we first investigate its asymptotic limits.

To this end, we turn the Langevin equation (Eq.~\ref{eq:activeevol}) 
into the corresponding Fokker-Plank equation 
for the probability distribution $P(q,v,t)$, 
where $v = \dot{q}$ is the velocity:
\begin{align}
    \label{eq:FP}
    \pdv{P}{t} + v\,\pdv{P}{q} + \frac{F(q)}{\gamma\tauA}\,\pdv{P}{v} = 
    \frac{\AA}{\tauA^2}\,\pdv[2]{P}{v} 
    + \frac{\Gamma(q)}{\gamma \tauA} \pdv{}{v}\l(vP\r)~.
\end{align}
For an active noise with long correlation time, $\tauA \gg \tau_0$ and strength $\lang \FH^2 \rang \ll \Fmax^2$, 
\begin{equation}
\label{eq:slow}
\kapslow  \simeq \l(2\tauA \r)^{-1}
\exp\l(-\half\Fmax^2 /\lang\FH^2\rang \r)~.
\end{equation}
This behaviour is specific to the active noise realisation which relies on $\Fmax$ and is markedly different from purely thermal reaction rate which depends on $\EB$. The relevance of $\Fmax$ in the long memory regime~($\tauA \gg \tau_0$) has been noted in a few recent studies~\cite{Woillez2019, Woillez2020, Woillez2020a} which did not consider hydrodynamic coupling. For the case of short memory, $\tauA \ll \tau_0$ and $\AA \ll \gamma\EB$, the active noise merely scales the energy barrier and the reaction rate follows the well-known thermal behaviour, 
\begin{equation}
\label{eq:fast}
\kapfast \simeq \l(\sqrt{2}\pi\tau_0 \r)^{-1} 
\exp\l[-\gamma\EB/\AA \r]~.
\end{equation}
We note that while $\kapfast$ grows with $\AA$, \ie with $\tauA$ for a fixed noise strength, $\kapslow$ decays monotonically with $\tauA$, suggesting an intermediate time scale where $\kappa$ is optimal.
Numerical simulations of the reaction dynamics for pure active noise confirm this behavior (Fig~\ref{fig:Act_Asy}).

It is important to realise that these semi-analytic Eqs.~[\ref{eq:slow},\ref{eq:fast}] are only valid at the asymptotic limits of the purely active case. For the case of enzymes, the thermal fluctuations can compete with or even dominate the active forces, and are thus no longer negligible. Also, both conditions are impractical for enzymatic solutions, as the relevant correlation time is typically distributed within a wide bandwidth, much beyond these asymptotic limits. Therefore, as these regimes cannot be accessed analytically, we now move on to solve Eqs.~[\ref{eq:eom},\ref{eq:activeevol}] numerically to investigate the effect of both active and thermal noise on the reaction rate.
We measure the simulation time and length in units of $\tau_0$ and $q_0$, respectively. Thermal and active fluctuations are also scaled by the relevant force scale as: $\sigT^2=\kBT/(2 \EB)$, $\sigA^2=\lang\FH^2\rang/(4a \EB)$~(See Method for Details). In the next section, we present behaviour of $\kappa$ in the $\{\sigA,\tauA\}$ plane spanning over orders of magnitude. Specifically, for each point in the $\{\sigA,\tauA\}$ plane, we generate $\num{e5}$ independent reaction trajectories starting from an initial position chosen randomly around $q_0$ and evolve the trajectories till it crosses $q=0$ where the energy barrier is maximum. The reaction rate $\kappa$ is then computed as the inverse of the mean time taken by trajectories to cross the barrier. 
Though this criterion of choosing a crossing point is not unique, the other choices of crossing points right to $q=0$, \ie allowing barrier recrossing, only result in insignificant changes in the absolute values of the mean first passage times. The results presented below are thus not dependent on such choices.

\begin{figure*}[!htb]
\centering
\includegraphics[width=0.93\textwidth]{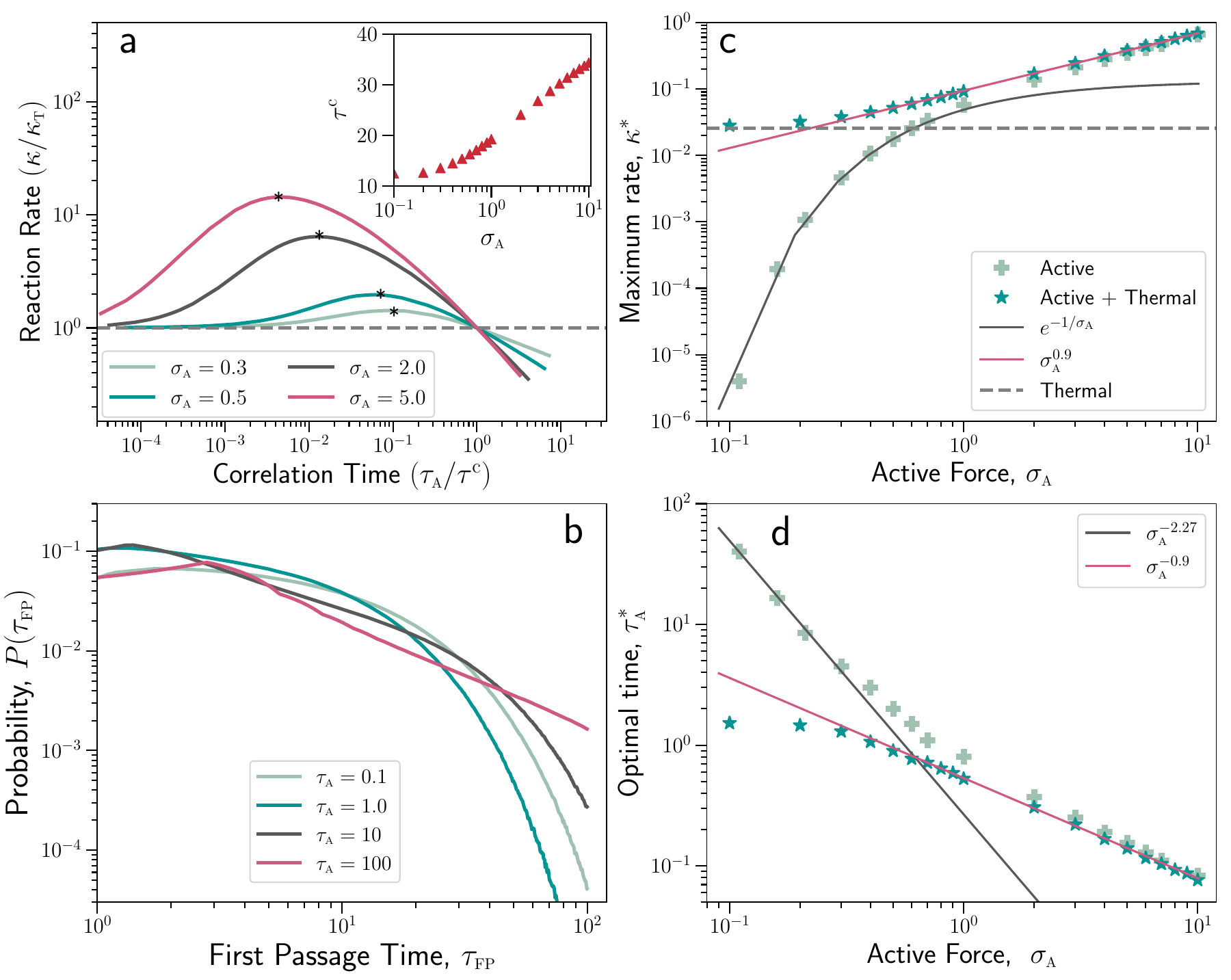}
\caption{\sb{Non-monotonic enhancement of reaction rate in the presence of active noise.}
(\sb{a}) The effect of active noise strength $\sigA$ on reaction rate $\kappa$ for a energy barrier $\EB=2\kBT$. The variation of $\kappa$ relative to thermal reaction rate $\kappaT$ for different values of $\sigA$ is plotted as a function of $\tauA$ scaled by $\tauc$, where $\tauc$ is the time above which the reaction rate is slower than the thermal rate. The inset shows the dependence of $\tauc$ on $\sigA$. For each value of $\sigA$, the maximum reaction rate $\kappa^*$ is pointed by an asterisk. 
(\sb{b}) Distribution of first passage time $P(\tauFP)$, exponential for small $\tauA$, starts to grow a prominent power-law tail with increasing $\tauA$ indicating departure from equilibrium. 
(\sb{c}) The maximal rate $\kappa^*$ (stars) is never less than $\kappaT$ (dashed line) and increases monotonically with the active noise $\sigA$. Also shown are the rates in a purely active system ( crosses). 
(\sb{d}) The optimal correlation time $\tauA^*$ associated with the maximal rate $\kappa^*$ (stars) shows a monotonic decrease with noise strength $\sigA$. Also shown is the purely active case (crosses). The numerical fits in (\sb{c}) and (\sb{d}) (black and red lines) are empirical functions described in the legends.
}
\label{fig:NS-AF}
\end{figure*}
\nsb{Enhancement of reaction rate by active noise}
Active noise changes the magnitude of the overall force that a reaction experiences in a given reaction energy landscape. It also changes the persistence of the force direction by introducing a correlation timescale that is absent in purely thermal agitation. As a result of this combined effect, the reaction rate $\kappa$ is expected to change. To investigate, we consider a case where thermal and active noises have equal strength: $\sigA=\sigT=0.5$, and plot $P(q)$, the probability distribution of the reaction trajectories in the reaction landscape, for different values of $\tauA$ along with the purely thermal case.~(Fig.~\ref{fig:probability}\sb{a}) Note that the probability of forming reactant-substrate activated complex, $P(q=0)$, is larger than the thermal case for $\tauA\sim\tau_0$ but becomes smaller for $\tauA=10\tau_0$. Following the position of the most probable value $\qmax$ as a function of increasing $\tauA$, we find that $\qmax$ gradually moves outwards from its thermal equilibrium position $\qmax=q_0$~(Fig.~\ref{fig:probability}\sb{b}). The movement is more rapid  over an intermediate range of $\tauA$ while for small $\tauA$, $\qmax$ mostly stays close to $q_0$ and at large $\tauA$, it somewhat settles at a certain value of $q$. However, the maximum value of the probability $\Pmax$ drops below its thermal value even when an active noise with a tiny correlation is introduced (Fig.~\ref{fig:probability}\sb{c}). $\Pmax$ continues to decrease for $\tauA \sim \tau_0$. As the correlation time of the active noise grows longer than the thermal crossing time, the memory of active noise starts to affect the reaction adversely. The reaction trajectories now stay away from the barrier for a longer time causing $\Pmax$ to increase. Eventually, $\Pmax$ becomes larger than its pure thermal counterpart for active noise with $\tauA \gtrsim 10 \, \tau_0$. 

The variation of $\Pmax$ indicates that the active noise affects the reaction by effectively modifying the energy barrier. To confirm this, we construct an effective potential from the probability distribution as: $V(q) =-\ln{[P(q)/P(0)]}$, and compare the effective barrier $\Ebeff \equiv V(0)-V(\qmax)$ with the thermal barrier. A decrease in $\Ebeff (< \EB)$ is clearly observed for small and intermediate $\tauA$~(Fig.~\ref{fig:probability}\sb{d}), where enhancement of $\kappa$ is naturally expected. While $\Ebeff$ traces the same non-monotonic behavior of $\Pmax$, it helps us to identify a crossover timescale $\tauc$ above which $\Ebeff >\EB$, and the reaction becomes even slower than a pure thermal case. Thus, $\tauc$ provides us a natural threshold to discern between {\em fast} ($\tauA<\tauc$) and {\em slow} ($\tauA>\tauc$) active noise, \ie the background hydrodynamic fluctuations. 
Similar patterns are observed for a larger activation barrier, $\EB=10\kBT$, with the lowest $\Ebeff$  at $\tauA\approx 50\tau_0$ and crossover at $\tauc\approx 10^5\tau_0$~(Fig.~\ref{fig:probability}\sb{e}-\sb{g}), thereby confirming that the non-monotonic enhancement of the reaction by active fluctuations is a general hallmark.

Next, we examine the effect of the active noise strength $\sigA$ on the reaction rate $\kappa$. We find that $\sigA$ enhances the effect of $\tauA$ on $\kappa$ as we plot it relative to the thermal reaction rate $\kappaT$ as a function of the scaled correlation time $\tauA/\tauc$~(Fig.~\ref{fig:NS-AF}\sb{a}). Evidently, $\kappa$ becomes faster against the fast background and slower against the slow background. Note that $\tauc$ increases with $\sigA$ (Fig.~\ref{fig:NS-AF}\sb{a} Inset), demonstrating that larger $\sigA$ allows a longer window of $\tauA$ for reaction rate enhancement.

\begin{figure*}[!htb]
\includegraphics[width=\textwidth]{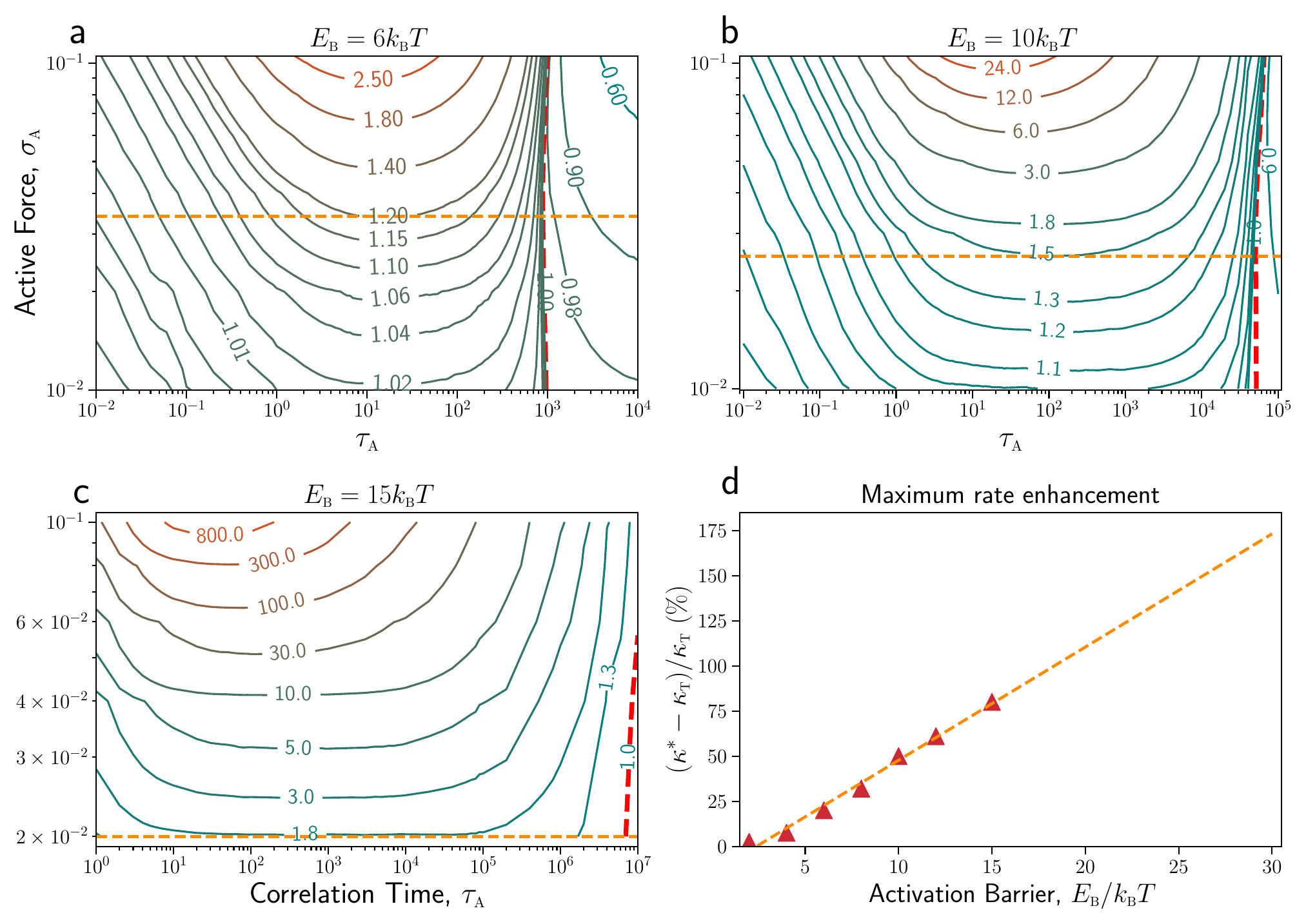}
\caption{
\sb{The effect of active noise on the catalytic rate of enzymes.} Variation of reaction rate $\kappa$, relative to thermal reaction rate $\kappaT$, as a function of active noise strength $\sigA$ and its correlation time $\tauA$, for  activation barriers, (\sb{a}) $\EB= \SI{6}{\kBT}$~($\kappaT = 10^{-3}$), (\sb{b}) $\EB= \SI{10}{\kBT}$~($\kappaT = \num{2e-5}$), and (\sb{c}) $\EB= \SI{15}{\kBT}$~($\kappaT = \num{1.23e-7}$). The solid curves represent the constant reaction rate contours with the rate increasing from blue to red contours.  The red dashed contour marks the line where $\kappa=\kappaT$, even in the presence of active fluctuations. This line separates the slow (right) and the fast (left) background regimes. Fast background always promotes enhancement of $\kappa$, which can be quite substantial depending on the control parameters. In contrast, slow background somewhat decelerates the reaction but not as spectacularly as in the fast background regime. For enzymatic solutions of concentration \SIrange{30}{300}{\micro\Molar}, $\sigA$ varies between \numrange{0.003}{0.034}, with a maximum value, denoted by orange dashed line. Within this range, a maximum enhancement of $\kappa$ up to \SI{20}{\percent} is expected for (\sb{a}) $\EB = \SI{6}{\kBT}$, up to \SI{50}{\percent} for (\sb{b}) $\EB = \SI{10}{\kBT}$, and \SI{80}{\percent} for (\sb{c}) $\EB = \SI{15}{\kBT}$. (\sb{d}) The variation of the maximum enhancement in the reaction rate (red triangles) as a function of activation barrier $\EB/\kBT$. The dashed orange line shows a linear fit, $-15+6.3\cdot\EB/\kBT$, to the data. The rate enhancement at higher activation barrier can be extrapolated from this linear behavior.
}
\label{fig:NFh}
\end{figure*}

Most importantly, it is possible to find an optimal correlation time $\tauA^*$ for which the enhancement of reaction rate is maximum. This maximum reaction rate $\kappa^*$ is denoted by an asterisk for each $\sigA$ value in Fig.~\ref{fig:NS-AF}\sb{a}. Notice that the enhancement of $\kappa$ is possible for even a tiny value of $\sigA$ (Fig.~\ref{fig:NS-AF}\sb{c}), but that would require a relatively larger optimal correlation time $\tauA^*$~(Fig.~\ref{fig:NS-AF}\sb{d}, also see Fig.~\ref{fig:NFh}). Still, $\tauA^*$ is smaller than $\tauc$ by at least one order-of-magnitude, as rate enhancement can only occur in the presence of a fast hydrodynamic background. More enhancement is observed with increasing $\sigA$ as $\kappa^*$ grows in a scale-free fashion with $\sigA >1$. Correspondingly, $\tauA^*$ decreases in a similar fashion. As an aside, we mention that similar behavior is also expected when the active noise is much stronger than the thermal one and solely dictates the reaction. In this case, $\kappa^*$ would decrease exponentially for small $\sigA < 1$, markedly different than the more realistic scenario of enzymatic catalysis governed by both thermal and active noise.

The non-monotonic behavior of $\kappa$ (Fig.~\ref{fig:NS-AF}\sb{a}) is the outcome of the interplay of two competing effects. First, the reaction dynamics change as fluctuations cross over from a fast to an adiabatic regime.  To understand this effect, note that reaction dynamics in the presence of active noise can be considered as motion within a fluctuating reaction energy landscape, $\Ueff(q,t)=U(q)-q\zA(t)$, with a fluctuating effective energy barrier $\EBeff(t) \simeq \EB + q_0 \zA(t)$ (akin to Bell's law~\cite{Bell1978}). The persistence of the fluctuations is controlled by correlation time $\tauA$. When the fluctuations of the landscape are much faster than the enzymatic timescale, $\tauA \ll \tau_0$, the enzyme experiences an average effective barrier, $\langle \EBeff \rangle = \EB$, and the resulting rate is $\kappa \sim \exp(-\langle \EBeff \rangle) = \exp(-\EB) \sim \kappaT$, \ie close to the thermal rate. But when the fluctuations become more persistent, they approach an adiabatic regime, $\tau_0 <\tauA < 1/\kappaT$, where each crossing event occurs in a practically static potential and effective barrier. In this regime, the average rate will be the average over the static potentials,  $\kappa \sim \langle \exp(-\EBeff) \rangle \ge  \exp(-\langle \EBeff \rangle)$, which is always larger than the rate in the fast regime, thus explaining the increasing part of the curve. 
This follows from the convexity of the logarithm (Jensen's inequality) \\
$\lang -\EBeff\rang = \lang\log\l[ \exp(-\EBeff) \r] \rang\le \log\lang \exp(- \EBeff)\rang $.
The second effect occurs in the large correlation limit, when $\kappa$ is controlled by the maximum force, $\Fmax$ rather than the activation barrier. Then, the reaction rate $\kappa$ exhibits an inverse dependence on $\tauA$~(Eq.~[\ref{eq:slow}]), due to slowing down by the increasing effective friction, $\Gamma(q) \sim \tauA$ in Eq.~[\ref{eq:eom2}]. 

Interpolating these two limits, one expects an optimal correlation time, where the reaction rate attains a maximum as indeed shown in the simulations. These observations agree with the computed distribution of first passage time, $P(\tauFP)$, the time taken to cross the reaction barrier (Fig.~\ref{fig:NS-AF}\sb{b}). The $P(\tauFP)$ distribution follows a  non-monotonic dependence similar to that of $\tauMFP$. For $\tauA\sim\tau_0$, the $P(\tauFP)$ shifts to shorter $\tauFP$ values compared to the thermal regime ($\tauA \ll \tau_0$), resulting in increasing reaction rate. On the other hand, for $\tauA \gg \tau_0$, the distribution shifts toward the longer first passage times, indicating slowing down compared to the thermal rate $\kappaT$. 
$P(\tauFP)$ crosses over from exponential scaling in the fast regime to power-law behaviour in the slow regime, signaling a transition from equilibrium to nonequilibrium behavior.

\begin{figure}[!htb]
\includegraphics[trim = 1cm 0cm 2.5cm 0.3cm,clip, width=0.95\columnwidth]{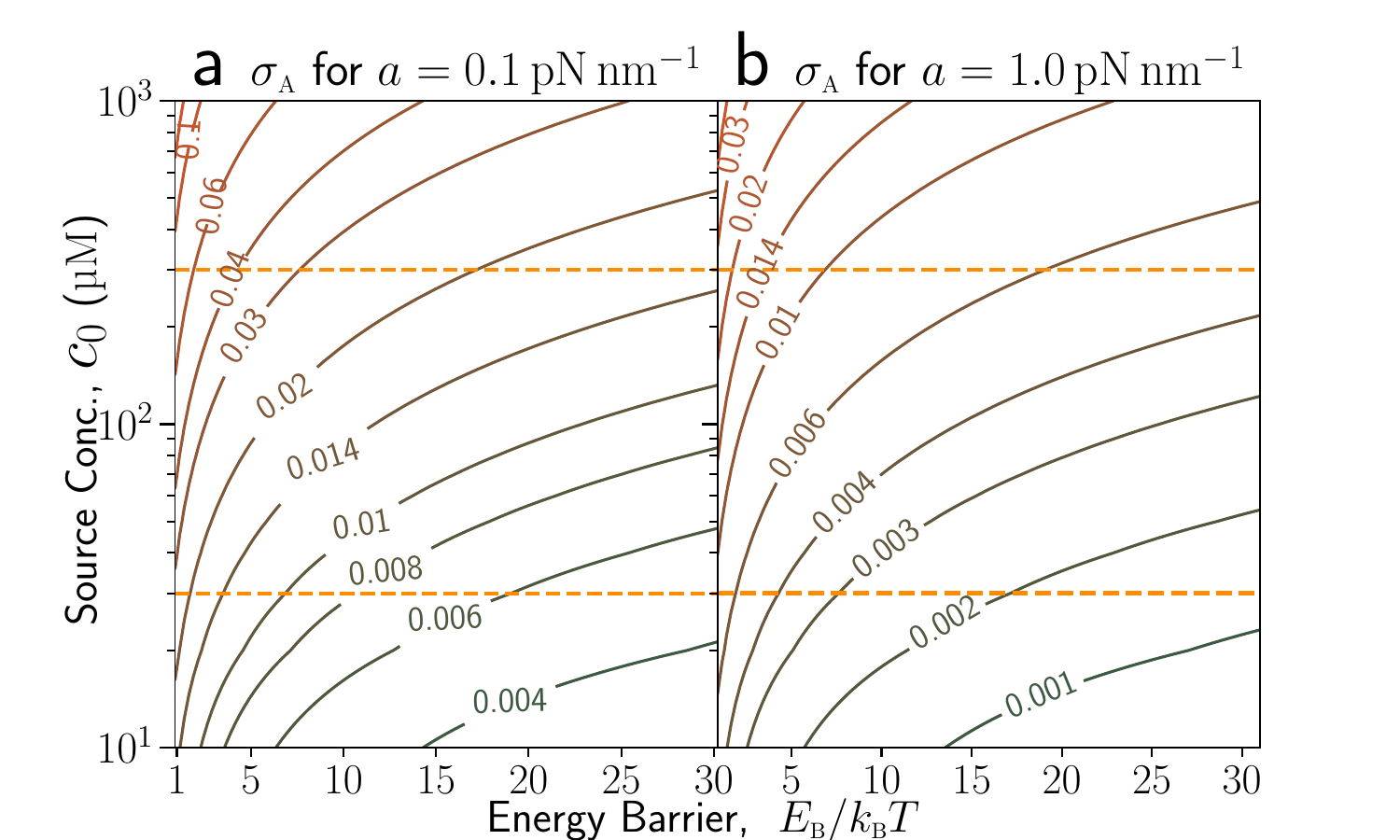}
\caption{\sb{Variation of scaled active force strength as a function of source dipole concentration and activation barrier.}
 (\sb{a}) and (\sb{b}) correspond to the protein stiffness values $a=0.1$ and $\SI{1.0}{\pico\newton\per\nm}$, respectively. Solid curves are constant force contours, and the dashed lines denote the concentration range, \SIrange{30}{300}{\micro\Molar}, which can be accessible in the experiments.
} 
\label{fig:ActF}
\end{figure}

\nsb{The case of enzymes}
Finally, we examine enzymatic catalysis in the presence of an actively fluctuating hydrodynamic background. During catalysis, structural and energetic transitions in the enzyme occur in multiple steps: typically, starting from the closure of a specific domain upon substrate binding, followed by the chemical step consisting of the chemical reaction and product release, and re-opening of the binding domain. Such a multi-step sequence has been observed, for example, in adenylate kinase (ADK) in which reopening of the bounded domain is identified as the rate-limiting step.~\cite{WolfWatz2004,HenzlerWildman2007}

To assess the effect of active hydrodynamic fluctuations on  catalytic cycles, we evaluate the change in reaction rate as a function of the relevant active noise parameters, \{$\sigA$, $\tauA$\}.
The scaled active noise strength, $\sigA=\sqrt{\langle\FH^2\rangle/(4a\EB)}$, depends on the reaction energy barrier $\EB$, enzyme stiffness $a$, and the density of the background through $\langle \FH^2 \rangle$ as in Eq.~(\ref{eq:hydroforce}). Catalytic reaction energy barriers are measured to typically lie within a range of $\EB=\SIrange{4}{30}{\kBT}$~\cite{Fersht2017, Liu2013} and the typical stiffness of enzymes is reported to vary within a range $a = \SIrange{0.1}{1.0}{\pico\newton\per\nm}$.~\cite{Shiroguchi2011, Lewalle2008, Alemany2016, Hosaka2020a} 
In Fig.~\ref{fig:ActF}, we have charted out the variation of $\sigA$ over a wide range of backgrounds with density ranging between \SIrange{10}{1000}{\micro\Molar} as a function of $\EB$ for two limiting values of $a$.

Now we compute the reaction rate for our test enzyme over a range of $\sigA$ and $\tauA$ for three different energy barriers, $\EB=\SI{6}{\kBT}$ (Fig.~\ref{fig:NFh}\sb{a}), $\EB=\SI{10}{\kBT}$~(Fig.~\ref{fig:NFh}\sb{b}), and $\EB=\SI{15}{\kBT}$~(Fig.~\ref{fig:NFh}\sb{c}). A red dashed line is drawn to mark the boundary between the fast background (on the left) and slow background (on the right). Within the fast regime, we always find an enhancement over the thermal reaction rate, $\kappa/\kappaT>1$. As a crowded solution of dipoles would correspond to a concentration, $(1/\ell_0^3)\simeq\SI{10}{\milli\Molar}$, we consider a moderate regime of \SIrange{30}{300}{\micro\Molar}. Using the active force map in Fig.~\ref{fig:ActF}, we find that an enzyme is expected to experience a maximal active force, $\sigA \simeq \numrange{0.003}{0.034}$ (shown as a horizontal dashed line in Fig.~\ref{fig:NFh}) over the relevant energy scale, $\EB = \SIrange{6}{15}{\kBT}$. At this limit, we see a maximum enhancement of $\SI{20}{\percent}$ for $\EB=\SI{6}{\kBT}$ ($\kappa/\kappaT \simeq 1.2$), of $\SI{50}{\percent}$ for $\EB=\SI{10}{\kBT}$ ($\kappa/\kappaT \simeq 1.5$), and of $\SI{80}{\percent}$ for $\EB=\SI{15}{\kBT}$ ($\kappa/\kappaT \simeq 1.8$).~(Fig.~\ref{fig:NFh})

Note that as $\EB$ increases, the barrier crossing events become exponentially rare. Thus, accessing the relevant range of $\sigA$ becomes very expensive computationally. For example, the activation barrier for ADK reported to be $\EB\approx \SI{25}{ \kBT}$.~\cite{WolfWatz2004,Arora2007,Kerns2015} Such barrier would correspond to thermal mean first passage time, $\tauMFP = \num{1.8e11}\tau_0$, which is at least four orders of magnitude longer than the case of $\EB=\SI{15}{\kBT}$, the largest strength barrier reported above, rendering the simulation too slow to be practical. Nevertheless, the maximum enhancement in the reaction rate $\kappa^{*}$ is well approximated by a linear function of the activation barrier $\EB$ (Fig.~\ref{fig:NFh}\sb{d}). Extrapolating this curve, we would expect an increase of roughly $\SI{140}{\percent}$ in the catalytic reaction rate of ADK at $\EB \approx \SI{25}{\kBT}$, while the $\SI{80}{\percent}$ increase observed at $\EB \approx \SI{15}{\kBT}$ serves as a lower bound.

\vspace{0.8cm}
\noindent\sb{\large Conclusion}\\
In summary, we have shown how a fluctuating hydrodynamic background might affect enzymatic catalysis. Hydrodynamic fluctuations of various origins are considered as an outcome of the stochastic oscillations of a random distribution of force dipoles. Coupled through the flow they generate, these force dipoles can be collectively realized as a temporally-correlated athermal noise representing the background activity. Modeling active noise as an Ornstein-Uhlenbeck process and numerically solving reaction rate theory, now in presence of both thermal and active noise, reveals a special correlation time $\tauc$, above which reaction rate start to slow down compared to the bare thermal rate. $\tauc$ is of the same order as the inverse of thermal reaction rate $\kappaT$~(red dashed line in Fig.~\ref{fig:NFh}). Further, we find that while a slow background, $\tauc < \tauA$, somewhat slows down the catalytic activity, a faster background, $\tauA<\tauc$, always enhances the catalytic reaction rate relative to the purely thermal case. For example, a physically realizable value of active noise may result in up to $\SI{140}{\percent}$ enhancement for the typical example of ADK. We note that the present model assumes Oseen's far-field approximation for the mobility tensor, and should be modified for densely packed sources. Once the hydrodynamic interaction is corrected to account for near-field effects, our dynamical equations can be solved in this limit of intense active force.

The proposed physical scenario and the predicted effect of active noise on enzymatic catalysis require cautious examination. As controlling the background is hard in vivo, we propose a simple in vitro experimental test: Consider a solution consisting of two enzymes and their respective substrates in an appropriate buffer. Importantly, the two reactions are chemically orthogonal to avoid any cross-talk. The ``source" enzymes generating the active noise are relatively dense to allow a strong impact on the ``target" enzymes, which are diluted to avoid confounding inverse effects. The active noise correlation time depends on the conformational step of the reaction of the source enzyme. Thus the correlation time is smaller than the inverse of thermal reaction rate of source enzyme, $\tauA < 1/\kappaT^{\source}$. Also, notice that the crossover time $\tauc$ remains of the order of the thermal $\tauMFP$ of the target enzyme, $\tauc \simeq 1/\kappaT^{\target}$~(red dashed line in Fig.~\ref{fig:NFh}). Thus, $\kappaT^{\source} > \kappaT^{\target}$, implies that the source serves as fast background , $\tauA < \tauc$, and we expect an enhancement in the reaction rate of target enzyme. Conversely, $\kappaT^{\source} < \kappaT^{\target}$, implies a source that serves as a slow background, $\tauA>\tauc$, and is expected to slow down the rate of the target enzyme. Thus, in general, a  separation of timescales between the target enzyme ($\tauc$) and the active noise ($\tauA$) is required to obtain a measurable effect on the reaction rate of target enzymes. 

As a consequence, in a solution of only one enzyme, which serves as both target and source, the enhancement will depend on the time scale of the conformational motion, which includes the closing and opening steps of enzyme domains. Generically, one of these steps is faster than the thermal reaction time $\tauA < 1/\kappaT$ (which is determined by another rate-limiting step), and we therefore anticipate a measurable self-enhancement of enzymatic reaction.
In the same spirit, the present model should be also applicable to other biologically relevant processes such as unzipping of DNA hairpins~\cite{Woodside2006,Woodside2008,Greenleaf2008} for which the reported energy~\cite{Neupane2012} and timescales \cite{Vandebroek2017} lie within the range explored in the current study. We note that the proposed rate-enhancement mechanism demonstrated here for enzymatic catalysts~\cite{Jee2018,Jee2018a} is general and may apply also to smaller organic catalysts~\cite{Wang2020} where recent experimental evidence indicates boosted mobility and long-range hydrodynamic interactions.

The exact nature of the mechano-chemical coupling during the catalytic cycle is an open question, a matter of active debate. The reaction energy landscape is often very complex and multidimensional, with numerous possibilities for energy exchange and conformational changes.~\cite{Benkovic2008,Kerns2015,Li2015} Nevertheless, the overall turnover rate typically depends on the rate-limiting step of crossing the highest energy barrier. In the vicinity of the crossing, the landscape is effectively one-dimensional, and the methodology developed here is therefore applicable to scenarios where the rate-limiting step is governed by conformational dynamics. Whether mechanical deformation can also affect the reaction rate when chemical steps are rate-limiting remains unclear. Recent evidence suggests that the conformational motion modulates the electric field by altering the position of the residues in the active site,~\cite{Welborn2019,Fried2017,Zoi2017} thus potentially affecting the chemical step. We plan to extend the present framework to address such scenarios.

The complex cellular environment is dense in entangled energetic processes. A fast-growing bacterium consumes energy at a power of ${\sim}\SI{e8}{\kBT\per\s}$, over a volume of ${\sim}\SI{1}{\um^3}$.~\cite{Milo2015} In the eukaryotic cell, there are high-activity regions and organelles, such as mitochondria and chloroplasts, where the proposed effects might be significant. One may speculate that molecular motors, whose turnover rate is relatively slow,~\cite{Howard1997,Milo2015} can be accelerated in the presence of high metabolic activity. To treat such elaborate scenarios, we plan to further extend the present bare-bone model to include the causal dependence of reactions in a network and the spatiotemporal heterogeneity of the embedding background. We hope the current results would stimulate further study of the potential effects of an active stochastic environment on biochemical processes.

\appendix

\vspace{0.5cm}
\noindent\sb{Methods}\\
{\smaller
\nsb{Hydrodynamic forces induced by active processes}
We consider a solution of stochastic force dipoles ~\cite{Mikhailov2015, Hosaka2020} 
representing active processes such as enzymatic catalysis and the motion of molecular motors. In this coarse-grained view, each force dipole consists of two beads connected by a spring of equilibrium length $\ell_0$. The beads represent the domains of the enzyme that move with respect to each other during the catalysis. 
Consider a collection of force dipoles \{$\mb_i$\} located at positions $\{\Rb_i\}$ with random independent orientations $\{\eb_i\}$ (3D unit vectors). Our target dipole has its two domains (\ie spheres) located at positions $\Rb$ and $\Rb'=\Rb + \ell \eb$, with an orientation $\eb$ and distance $\ell = |\Rb'- \Rb|$. Following Mikhailov and Kapral,~\cite{Mikhailov2015} we find the velocities $\dot{\Rb}$ of the domains---using the mobility tensor $\GG_{\alpha\beta}$---by summing the contributions of the velocity fields 
induced by the surrounding dipoles,
\begin{align}
\dot{R}_{\alpha} &= \sum_i^N 
\pdv{\GG_{\alpha\beta}\l( \Rb-\Rb_i \r)}{R_{i\mu}}
 e_{i\beta}e_{i\mu}m_i(t)~,\\
\dot{R'}_{\alpha} &= \sum_i^N
\pdv{\GG_{\alpha\beta}\l(\Rb'-\Rb_i \r)}{R_{i\mu}} 
e_{i\beta}e_{i\mu}m_i(t)~,
\label{eq:DipVel}
\end{align} 
where the Greek indices denote $x,y,z$ components of vectors and tensors, and we follow Einstein's convention of summation over repeated indices. The mobility tensor $\GG_{\alpha\beta}$ is the Green function of the linear Stokes flow, which yields the velocity field resulting from a localized force.~\cite{happel1983} Eq.~[\ref{eq:DipVel}] involves spatial derivatives of $\GG$, which are the Taylor expansions around each force dipole. 

The time-dependent dipole moment exerted on the target $m(t)$ is $m(t)=\ell(t)F(t)$, where $\ell(t)$ and $F(t)$ are the distance and interaction force between the two domains. Thus, for a target enzyme of length $\ell(t)$, the relative velocity $\delta \vb$ between the two domains is given by
\begin{align*}
\delta v_{\alpha} &= \dot{R'}_{\alpha} -\dot{R}_{\alpha}= \sum_i^N e_{i\beta}e_{i\mu}m_i(t)\nonumber \\
&\l[ \pdv{\GG_{\alpha\beta}\l( \Rb+\ell(t)\eb-\Rb_i \r)}{R_{i\mu}}-
\pdv{\GG_{\alpha\beta}\l(\Rb-\Rb_i \r)}{R_{i\mu}}\r]~.
\end{align*}
Since the linear extension of the enzyme $\ell(t)$ is much smaller than the dipole-dipole distances, we can take a far-field approximation by expanding the difference to first order in $\ell$, 
\begin{align*}
\delta v_{\alpha} = \ell(t) \sum_i
\pdv{\GG_{\alpha\beta}\l(\Rb-\Rb_i \r)}{R_{i\mu}}{R_{\nu}}
e_{i\beta}e_{i\mu}e_{\nu}m_i(t)~.
\end{align*} 
Therefore, the relative velocity with which the spring connecting the two domains compresses or stretches is the projection
\begin{align*}
\delta \vb \cdot\eb = \ell(t) \sum_i
\pdv{\GG_{\alpha\beta}(\Rb-\Rb_i)}{R_{i\mu}}{R_{\nu}}
e_{i\beta}e_{i\mu}e_{\nu}e_{\alpha} m_i(t)~.
\end{align*}
Applying Stoke's law, we find that the deformation forces acting on the target dipole is 
\begin{align}
\label{eq:HForce}
\FH(\Rb,t) &= \l( 3\pi\eta\ww \r) \l(\delta\vb\cdot\eb \r) 
=  3\pi\eta \ww \ell(t)\nonumber \\
    &\times\sum_i
    \pdv{\GG_{\alpha\beta}\l(\Rb-\Rb_i\r)}{R_{i\mu}}{R_{\nu}}
    e_{i\beta}e_{i\mu}e_{\nu}e_{\alpha}m_i(t)~,            
\end{align}
where $\ww$ is the domain size (\ie its hydrodynamic diameter) and $\eta$ the viscosity of the solution. Since enzymes are randomly oriented, then without any loss of generality, we take the target enzyme oriented along the $x$-axis, thereby simplifying Eq.~[\ref{eq:HForce}] into
\begin{equation}
\FH(\Rb,t) =  3\pi\eta \ww \ell(t)\sum_i
\pdv{\GG_{x\beta}(\Rb-\Rb_i)}{R_{i\mu}}{R_{x}}
e_{i\beta}e_{i\mu}m_i(t)~.
\label{eq:Fh}
\end{equation}

Next, we rewrite Eq.~[\ref{eq:Fh}] in a field-point notation, which will be convenient for further manipulation, 
\begin{align*}
\FH(\Rb,t) &=  3\pi\eta \ww \ell(t) \int \dd{\rb}~
\pdv{\GG_{x\beta}(\rb)}{r_{\mu}}{r_{x}}\nonumber \\
         \times  &\sum_i e_{i\beta}e_{i\mu}m_i(t)\,
         \delta\l(\rb-\Rb_i+\Rb\r).
\end{align*}
The mean force is proportional to the average over the sum of dipole moments, which vanishes due to the symmetry in the homogeneous isotropic solution,~\cite{Mikhailov2015} $\langle \FH(\Rb,t)\rangle \sim \langle m_i(t)\rangle = 0$. However, the second moment---that is the average squared force the target dipole experiences due to the collective fluctuations of other force dipoles---does not vanish,
\begin{align*}
&\lang \FH(\Rb,t) \FH(\Rb, t') \rang  =  
(3\pi\eta \ww)^2 \lang \ell(t)\ell(t') \rang \nonumber \\ 
& \times\int \dd{\rb}\,
  \pdv{\GG_{x\beta}(\rb)}{r_{\mu}}{r_{x}} \, 
  \pdv{\GG_{x\beta'}(\rb)}{r_{\mu'}}{r_{x}} \nonumber \\    
& \times \sum_i \lang e_{i\beta}  e_{i\mu}e_{i\beta '}e_{i\mu'}
\delta(\rb-\Rb_i+\Rb) m_i(t)m_i(t') \rang ,
\end{align*}
Since dipole orientations are uncorrelated with their positions, the last term in above equation can be simplified,
\begin{align*}
\sum_i &\lang e_{i\beta}e_{i\mu}e_{i\beta '}e_{i\mu '}
\delta(\rb-\Rb_i+\Rb) m_i(t)m_i(t') \rang  \nonumber \\
&=\lang e_{\beta}e_{\mu}e_{\beta '}e_{\mu '} \rang 
\lang m(t)m(t') \rang c\l(\Rb+\rb \r), 
\end{align*}
where $c(\rb) = \sum_i \delta(\rb-\Rb_i)$ is the concentration of force dipoles in the solution. Thus, we find that the second moment of the force is 
\begin{align*}
\langle \FH(\Rb,t)&\FH(\Rb, t')\rangle = 
\nonumber \l( 3\pi\eta \ww \r)^2 \lang \ell(t)\ell(t') \rang\\ 
 &\times\int \dd{\rb}\,
\pdv{\GG_{x\beta}(\rb)}{r_{\mu}}{r_{x}} 
\pdv{\GG_{x\beta'}(\rb)}{r_{\mu'}}{r_{x}}\,
c(\Rb+\rb) \nonumber \\    
&\times \lang e_{\beta}e_{\mu}e_{\beta '}e_{\mu'} \rang 
\lang m(t)m(t') \rang ~.
\end{align*}
Assuming a uniform concentration, the variance of this force is 
\begin{align}
\label{eq:VHForce}
\lang \FH^2(\Rb,t) \rang &= 
\l(3\pi\eta \ww \r)^2 \lang \ell^2(t) \rang  
\langle m^2(t)\rangle c_0 \nonumber \\
\times &\int \dd{\rb}\,
\pdv{\GG_{x\beta}(\rb)}{r_{\mu}}{r_{x}}
\pdv{\GG_{x\beta '}(\rb)}{r_{\mu'}}{r_{x}}
\lang e_{\beta}e_{\mu}e_{\beta '}e_{\mu '} \rang~.
\end{align}
Since dipolar orientation is uncorrelated, the $4$-point correlation term 
$\lang e_{\beta}e_{\mu}e_{\beta '}e_{\mu '} \rang$ vanishes unless there are even powers of the components of the orientation vector $\eb$. We can therefore write the $4$-point correlation as a sum over products of $\delta$-functions,
\begin{align*}
\lang e_{\beta}e_{\mu}e_{\beta '}e_{\mu '} \rang = 
A_d\l[ \delta_{\beta\beta '}\delta_{\mu\mu'}+
\delta_{\beta\mu}\delta_{\beta' \mu '}+\delta_{\beta\mu '}\delta_{\beta ' \mu}\r],
\end{align*}
where  $A_d=1/15$ for a 3D system.

To proceed further, we use a far-field approximation for $\GG_{\alpha\beta}$ in terms of the Oseen tensor,~\cite{Pozrikidis1992,happel1983} which for a 3D system is
\begin{equation}
\label{eq:G3d}
\GG_{\alpha\beta}(\rb) = 
\frac{1}{8\pi\eta r}
\l[\delta_{\alpha\beta}+\frac{r_{\alpha} r_{\beta}}{r^2} \r]~.
\end{equation}
The Oseen approximation is valid as long as the separation between dipoles is large compared to their size. Substituting Eq.~[\ref{eq:G3d}] in Eq.~[\ref{eq:VHForce}] and introducing a scaled coordinate $\xib=\rb/\ell_0$, we find
\begin{align}
\label{eq:VHForce3D}
\lang \FH^2(\Rb,t) \rang &=
\frac{9}{64} A_d\langle \ell^2(t)\rangle\langle m^2(t)\rangle 
c_0 \ww^2 \ell_0^{-3} \nonumber \\
 \times & \int_{1}^{\infty} \dd{\xib}
 \pdv{\GG_{x\beta}(\xib)}{\xi_{\mu}}{\xi_{x}} 
 \pdv{\GG_{x\beta '}(\xib)}{\xi_{\mu'}}{\xi_{x}} \nonumber \\
 \times & \l( \delta_{\beta\beta '}\delta_{\mu\mu'}+
 \delta_{\beta\mu}\delta_{\beta'\mu'}+
 \delta_{\beta\mu'}\delta_{\beta'\mu} \r) ~,
\end{align}
where the scaled Oseen tensor is $\GG_{\alpha\beta}(\xib) = \xi^{-1}(1+\xi_{\alpha}\xi_{\beta}/\xi^2)$. Since the mobility tensor diverges as $1/\xi^3$ at small distances, we introduce a cut-off in the lower limit of the integration accounting for the the finite size of the dipole (\ie~enzyme). The integral in the Eq.~[\ref{eq:VHForce3D}] is a dimensionless factor, which depends on the derivatives of $\GG_{\alpha\beta}(\xib)$ and the dipole orientations. A straightforward calculation yields
\begin{align*}
\int_{1}^{\infty} & \dd{\xib}
\pdv{\GG_{x\beta}(\xib)}{\xi_{\mu}}{\xi_{x}}
\pdv{\GG_{x\beta '}(\xib)}{\xi_{\mu '}}{\xi_{x}} \nonumber\\
\times & \l(\delta_{\beta\beta '}\delta_{\mu\mu '}+
\delta_{\beta\mu}\delta_{\beta ' \mu '}\nonumber  
+\delta_{\beta\mu '}\delta_{\beta ' \mu} \r)
  = \frac{96\pi}{5}~.
\end{align*}

Finally, substituting the value of integral in Eq.~[\ref{eq:VHForce3D}], we find the variance of the hydrodynamic force,
\begin{equation}
\label{eq:VHForce3Dc}
\lang \FH^2(\Rb,t) \rang=
\l( \frac{9\pi}{50} \cdot
\frac{\ww^2}{\ell_0^2} \cdot
\frac{\langle \ell^2\rangle}{\ell_0^2} \r)
\l(\frac{\lang m^2\rang}{\ell_0^2} \r)
\l( {c_0 \ell_0^3} \r) ~.
\end{equation}
The dependence of the hydrodynamic force on the inter-dipole distance $\RR$ arises from the dipole concentration $c_0 = 1/\RR^3$. The first three terms on the right-hand side of Eq.~[\ref{eq:VHForce3Dc}] are combined into a geometric factor $\lambda = (9\pi/50)\ww^2 \langle \ell^2\rangle/\ell_0^4$. This constant is of order $\lambda \simeq 1/2$ since all the three lengths are similar. We have used Eq.~[\ref{eq:VHForce3Dc}] to estimate the hydrodynamic force generated by an enzymatic solution.\\

\nsb{Barrier crossing under the combined influence of thermal and active noise} 
We examine overdamped Langevin dynamics in a reaction energy landscape $U(q)$ of a symmetric bistable system,
\begin{equation*}
    U(q) = -\frac{a}{2}q^2 +\frac{b}{4}q^4~.
\end{equation*}
$U(q)$ has two minima at $q_m = \pm \sqrt{a/b}$, separated by an energy barrier, $\EB = a^2/(4b)$. In the overdamped Langevin framework, the reaction coordinate $q$ evolves according to 
\begin{equation}
 \gamma\dot{q} = -\pdv{U}{q} + \zT(t) +\zA(t)~.
\label{eq:OLE}   
\end{equation}
The noise term $\zT(t)$ in Eq.[\ref{eq:OLE}] is a standard stochastic thermal force with the statistics 
\begin{equation*}
\langle \zT(t)\zT(t') \rangle 
= 2\gamma \kBT \delta(t-t')~,
\end{equation*} 
The active force $\zT(t)$ is modeled as an Ornstein-Uhlenbeck Process,
\begin{equation}
\label{eq:ANoise}
\tauA \dot{\zeta}_\sA = -\zA + \sqrt{2\AA}~\xiw(t)~,
\end{equation}
where $\xiw(t)$ is a white noise source with zero mean and unit variance, $\AA$ is the energy scale of the active force, and $\tauA$ correlation time of the activity. The corresponding active force statistics is given by
\begin{equation*}
\lang \zA(t) \zA(t') \rang 
= \l( \AA/\tauA \r)\,e^{-|t-t'|/\tauA}~.
\end{equation*} 
For an Ornstein-Uhlenbeck process, the fluctuation-dissipation relation implies that $\AA$ is proportional to $\tauA$. Hence, the variance of the active force, $\lang \FH^2 \rang=\langle \zA^2(t) \rangle = \AA/\tauA$, remains constant. 

To examine the impact of active noise on barrier crossing, we numerically solve many realizations of Eq.~[\ref{eq:OLE}] and analyze the statistics of crossing events.
For this purpose, we introduce the following scaling 
\begin{equation*}
\bt = t/\tau_0 ~,~\bq = q/q_0~,
\end{equation*} 
where $\tau_0 = \gamma/a$ is the thermal relaxation time of the particle in the vicinity of the minimum at $q_0$. Using the above scaling, we obtain a dimensionless form of the Eqs.~[\ref{eq:OLE},\ref{eq:ANoise}]
\begin{align}
\label{eq:OLES}
\dbq =&\, \bq-\bq^{~3} + \bzA\l(\bt \r) + \bzT\l(\bt \r)~,\\
\label{eq:ANoiseS}
\btauA\dot{\bzet}_\sA =&\, - \bzA + \sqrt{2\bAA}~\xiw\l(\bt \r)~,
\end{align} 
with $\bAA = \AA/(4\gamma \EB)$. The corresponding scaled noise statistics are 
\begin{align}
\label{eq:SNSSA}
\lang\bzT(\bt)\bzT(\bt') \rang &= 
\sigT^2\delta\l( \bt-\bt' \r), \nonumber\\
\lang \bzA(\bt)\bzA(\bt') \rang &=
\sigA^2\exp\l( -|\bt-\bt'|/\btauA \r),\nonumber\\
\lang\xiw(\bt)\xiw(\bt') \rang &= 
\delta\l( \bt-\bt' \r),
\end{align}
where $\sigA^2 = \lang \FH^2 \rang/(4a\EB)$, and $\sigT^2 = \kBT/(2 \EB)$  are the scaled active and thermal noise strength. Eqs.~[\ref{eq:OLES},\ref{eq:ANoiseS},\ref{eq:SNSSA}] 
are the central equations in our numerical and analytical study. To simplify the notation, we will hereafter omit the overbar in the scaled variables (so $q=\bq$ etc.).\\

\noindent
\textbf{The numerical simulation.~}
We solve Eqs.~[\ref{eq:OLES}] employing an explicit Euler scheme~\cite{Mannella2002}, which yields the following iterative dynamics for the reaction coordinate:
\begin{equation*}
q(t+dt) = q(t)(1+dt) -q^3(t)\,dt + X_{\zT} + X_{\zA}~,   
\end{equation*}
where $X_{\zT}$  and $X_{\zA}$ are the random processes
\begin{align}
   X_{\zT} = \int_{t}^{t+dt}\zT(u) \dd{u}~, \\
  \label{eq:XA} 
  X_{\zA} = \int_{t}^{t+dt}\zA(u) \dd{u}~.
\end{align}
The Gaussian distribution of the white thermal noise $\zT$ has zero mean, and a variance $\sigT^2$. Therefore, the distribution of $X_{\zT}$ is simply $X_{\zT} = \sqrt{dt}\,\sigT \YT$, where  $\YT \sim \NN(0,1)$ is distributed according to the standard normal distribution with zero mean and unit variance.

Integrating Eq.~[\ref{eq:ANoiseS}], we obtain a formal solution for the active noise,
\begin{equation*}
    \zA(t) = e^{-t/\tauA}\zA(0) + \frac{\sqrt{2\AA}}{\tauA}
    \int_0^{t}e^{(u-t)/\tauA}\xiw(u)\,\dd{u}~.
\end{equation*} 
Substitution of the latter result into Eq.~[\ref{eq:XA}], yields the statistics of $X_{\zA}$. To proceed further, we define two Gaussian processes,~\cite{Mannella2002,Sharma2017}
\begin{align*}
    \Omega_0 &= \int_0^{dt}{\dd{u} e^{(u-dt)/\tauA}\xiw(u)}~,\\
    \Omega_1 &= \int_0^{dt}{\dd{u}
                \int_{0}^{u} \dd{v} e^{(v-u)/ \tauA}\xiw(v)}~.
\end{align*}
Solving these equations, we express the $\Omega_0,\Omega_1$ processes as
\begin{align*}
    \Omega_0 &= \sqrt{\lang \Omega_0^2 \rang}~Y_0~, \\
    \Omega_1 &= \frac{\lang \Omega_0\Omega_1 \rang}
    {\sqrt{\lang\Omega_0^2 \rang}}~Y_0 + 
    \sqrt{\lang\Omega_1^2 \rang - 
    \frac{\lang\Omega_0\Omega_1 \rang^2}
    {\lang\Omega_0^2 \rang}}~Y_1~,
\end{align*}
with the correlations defined in terms of $\mu = dt/\tauA$ as 
\begin{align*}
    \lang\Omega_0^2 \rang &= \frac{\tauA}{2}\l(1-e^{-2\mu} \r)~,\\
    \lang\Omega_1^2 \rang &=
    \frac{\tauA^3}{2}\l(2\mu-3-e^{-2\mu}+4e^{-\mu} \r)~,\\
    \lang\Omega_0\Omega_1 \rang &=
    \frac{\tauA^2}{2}\l(1-2e^{-\mu}+e^{-2\mu} \r),
\end{align*}
and $Y_0 \sim \NN(0,1)$ and $Y_1 \sim \NN(0,1)$ are two independent standard Gaussian processes of zero mean and unit variance. With the expressions for the stochastic processes, the time update algorithm for active noise and reaction coordinate becomes
\begin{align*}
    \zA(t+dt) &= e^{-\mu}\zA(t) + \frac{\sqrt{2\AA}}{\tauA}~\Omega_0~,\\
    q(t+dt) &= q(t)(1+dt) -q^3(t)\,dt + \sqrt{dt}~\sigT \YT \nonumber \\ &+ \tauA \l( 1-e^{-\mu} \r)\zA(t) + \frac{\sqrt{2\AA}}{\tauA}~\Omega_1~. 
\end{align*}

To calculate the barrier crossing rate, we consider a particle, initially positioned at the left minimum $q=-1$ (\ie $q=-q_0$). We then monitor the particle trajectory and find the first passage time---the time when the particle crosses the energy barrier for the first time. We repeat the process for \num{e5} independent noise realizations and averaged to obtain mean first passage time $\tauMFP$. In a bistable system, the reaction rate $\kappa$ is inversely proportional to the mean first passage time,
$\kappa = \half\tauMFP^{-1}$. 
}\\

\bibliography{Ref-Act}

\vspace{0.8cm}

\nsb{Acknowledgements}
The research was supported by the Institute of Basic Science, South Korea, grant IBS-R020. We thank Ah-Young Jee and Steve Granick for essential discussions, and the anonymous referees for constructive comments. \\

\nsb{Data availability}
All the data presented in the paper are available from the authors upon reasonable request.\\

\nsb{Code availability}
Codes used for the simulation are available from the authors upon reasonable request.\\

\nsb{Author contributions}
All Authors -- AKT, TD, GP, HKT and TT -- contributed to the conceptualization of the project and were actively engaged in writing the manuscript. HKP and TT supervised the project. AKT and TT have done theoretical analysis and AKT performed the numerical simulations.\\

\nsb{Competing interest}
Authors declare no competing interest.
\end{document}